\documentclass[%
 a4paper, twocolumn,11pt
amsmath,amssymb
aps,floatfix,superscriptaddress,
]
{revtex4-1}
\pdfoutput = 1

\usepackage[utf8]{inputenc}
\usepackage[english]{babel}
\usepackage[T1]{fontenc}
\usepackage{xkeyval}
\usepackage{graphicx}
\usepackage[export]{adjustbox}
\usepackage{soul}
\usepackage{dcolumn}
\usepackage{bm}
\usepackage{epstopdf}
 \usepackage[usenames,dvipsnames]{pstricks}
 \usepackage{epsfig}
 \usepackage{pst-grad} 
 \usepackage{pst-plot} 
 \usepackage[space]{grffile} 
 \usepackage{etoolbox} 
 \makeatletter 
 \patchcmd\Gread@eps{\@inputcheck#1 }{\@inputcheck"#1"\relax}{}{}
 \makeatother
\usepackage{booktabs}
\usepackage{boxhandler}
\usepackage{amsmath}
\usepackage{subcaption}
\usepackage{braket}
\usepackage{amsmath}
\usepackage{float}
\makeatletter
\let\newfloat\newfloat@ltx
\makeatother
\usepackage{algorithm}
\usepackage{algpseudocode}

\usepackage{hyperref}
\hypersetup{
  colorlinks   = true,    
  urlcolor     = blue,    
  linkcolor    = blue,    
  citecolor    = blue      
}

\usepackage{multirow}
\usepackage{enumitem}

\newcommand{\Tr}{\mathrm{Tr}}
\newcommand{\Obs}{\mathcal{O}}
\newtheorem{definition}{Definition}

\begin{document}

\title{Comment on ``Recovering noise-free quantum observables''}

\author{Josu {Etxezarreta Martinez}}
\email{jetxezarreta@unav.es}
\affiliation{Department of Basic Sciences, Tecnun - University of Navarra, 20018 San Sebastian, Spain.}
\author{Olatz {Sanz Larrarte}}
\affiliation{Department of Basic Sciences, Tecnun - University of Navarra, 20018 San Sebastian, Spain.}
\author{Javier {Oliva del Moral}}
\affiliation{Department of Basic Sciences, Tecnun - University of Navarra, 20018 San Sebastian, Spain.}
\affiliation{Donostia International Physics Center, 20018 San Sebastian, Spain.}
\author{Reza {Dastbasteh}}
\affiliation{Department of Basic Sciences, Tecnun - University of Navarra, 20018 San Sebastian, Spain.}
\author{Ruben M. Otxoa}
\affiliation{Hitachi Cambridge Laboratory, J. J. Thomson Avenue, Cambridge, CB3 0HE, United Kingdom.}
\affiliation{Donostia International Physics Center, 20018 San Sebastian, Spain.}

\begin{abstract}
Zero-noise extrapolation (ZNE) stands as the most widespread quantum error mitigation technique in order to aim the recovery of noise-free expectation values of observables of interest by means of Noisy Intermediate-Scale Quantum (NISQ) machines. Recently, Otten and Gray proposed a multidimensional generalization of poynomial ZNE for systems where there is not a tunable global noise source [Phys. Rev. A \textbf{99,} 012338 (2019)]. Specifically, the authors refer to multiqubit systems where each of the qubits experiences several noise processes with different rates, i.e. a non-identically distributed noise model. The authors proposed a hypersurface method for mitigating such noise, which is technically correct. While effective, the proposed method presents an unbearable experiment repetition overhead, making it impractical, at least from the perspective of quantum computing. In this comment, we show that the traditional extrapolation techniques can be applied for such non-identically distributed noise setting consisted of many different noise sources, implying that the measurement overhead is reduced considerably. For doing so, we clarify what it is meant by a tunable global noise source in the context of ZNE, concept that we consider important to be clarified for a correct understanding about how and why these methods work.
\end{abstract}

\maketitle
\section{Introduction}
In a recent paper \cite{otten}, Otten and Gray proposed a zero-noise extrapolation (ZNE) method for multiqubit systems where each of the qubits experience multiple noise sources at different rates. Specifically, the Markovian noisy evolution in such setting can be described by the Lindblad master equation
\begin{align}\label{eq:inidLindblad}
    \frac{\partial}{\partial t} \rho(t) = -\frac{i}{\hbar}[H(t),\rho(t)] + \sum_{i} \lambda_i \mathcal{L}_i(\rho(t)),
\end{align}
where $H(t)$ is the Hamiltonian, each of the $\mathcal{L}_i(\rho(t))$ describes a possible noise source that the multiqubit system experiences with rate $\lambda_i$. Here, the dissipators $\mathcal{L}_i(\cdot)$ are generally consisted of jump operators, $L_i$, as $\mathcal{L}_i(\cdot) = L_i\cdot L_i^\dagger - \frac{1}{2}\{L_i^\dagger L_i,\cdot\}$, where $\{a,b\}=ab+ba$ is the anticommutator. The dissipators must be time-invariant and independent of the Hamiltonian drive \cite{TemmeZNE}.  For the analysis done here, we do not require to define the actual jump operators, but for the specific relaxation and dephasing noise considered in \cite{otten}, the system experiences $L_i^1=I\otimes\cdots\otimes\ket{0}\bra{1}_i\otimes\cdots\otimes I$ (relaxation) and $L_i^2=I\otimes\cdots \otimes Z_i\otimes\cdots\otimes I$ (dephasing \footnote{$Z$ refers to the Pauli Z matrix.}) jump operators with rates $\lambda_i^1$ and $\lambda_i^2$, respectively. This is an independent, non-identically distributed noise model, where each qubit experiences relaxation and dephasing independently and with different rates of interaction. These models integrate the non-uniformity observed in the noise of experimental quantum processors assuming independence, and have been also recently studied from the point of view of quantum error correction \cite{ton1,ton2}. Note, however, that equation \eqref{eq:inidLindblad} does not need to assume noise independence, as correlated terms can be plugged in, e.g. a 2 qubit dephasing term can be included as $I\otimes\cdots \otimes Z_i\otimes\cdots\otimes Z_j\otimes\cdots\otimes I$. Moreover, the dissipators in equation \eqref{eq:inidLindblad} can have other forms useful to represent other noise types, e.g. non-Markovian noise \cite{TemmeZNE}.

The authors in \cite{otten} discuss that such a noise model does not present one tunable global noise parameter, implying that the previously proposed ZNE techniques are not applicable \cite{TemmeZNE,ZNEexp1,ZNEcheby,ZNEtiron,QEMreview,QEMunify,Czarnik2021errormitigation}. Thus, the authors propose a method based on multivariate polynomial extrapolation in order to attempt to recover the ideal expectation value of interest. The main caveat in such protocol is the fact that the required measurement overhead increases to an impractical amount as a function of the noise sources and the multinomial approximation order. In this comment, we discuss how the requirement of ``one tunable global noise parameter'' is usually misinterpreted through the literature to then discus how standard ZNE methods \cite{TemmeZNE} can be applied to the setting discussed by Otten and Gray. We finish by showing how the overhead of the protocol is substantially reduced by means of such appropriate definition so that multiqubit systems with arbitrary noise sources and rates can be efficiently used to extrapolate ideal expectation values of observables. 

\section{Interpreting the meaning of one tunable global noise parameter}
\begin{figure}[h!]
    \centering
    \includegraphics[width=\columnwidth]{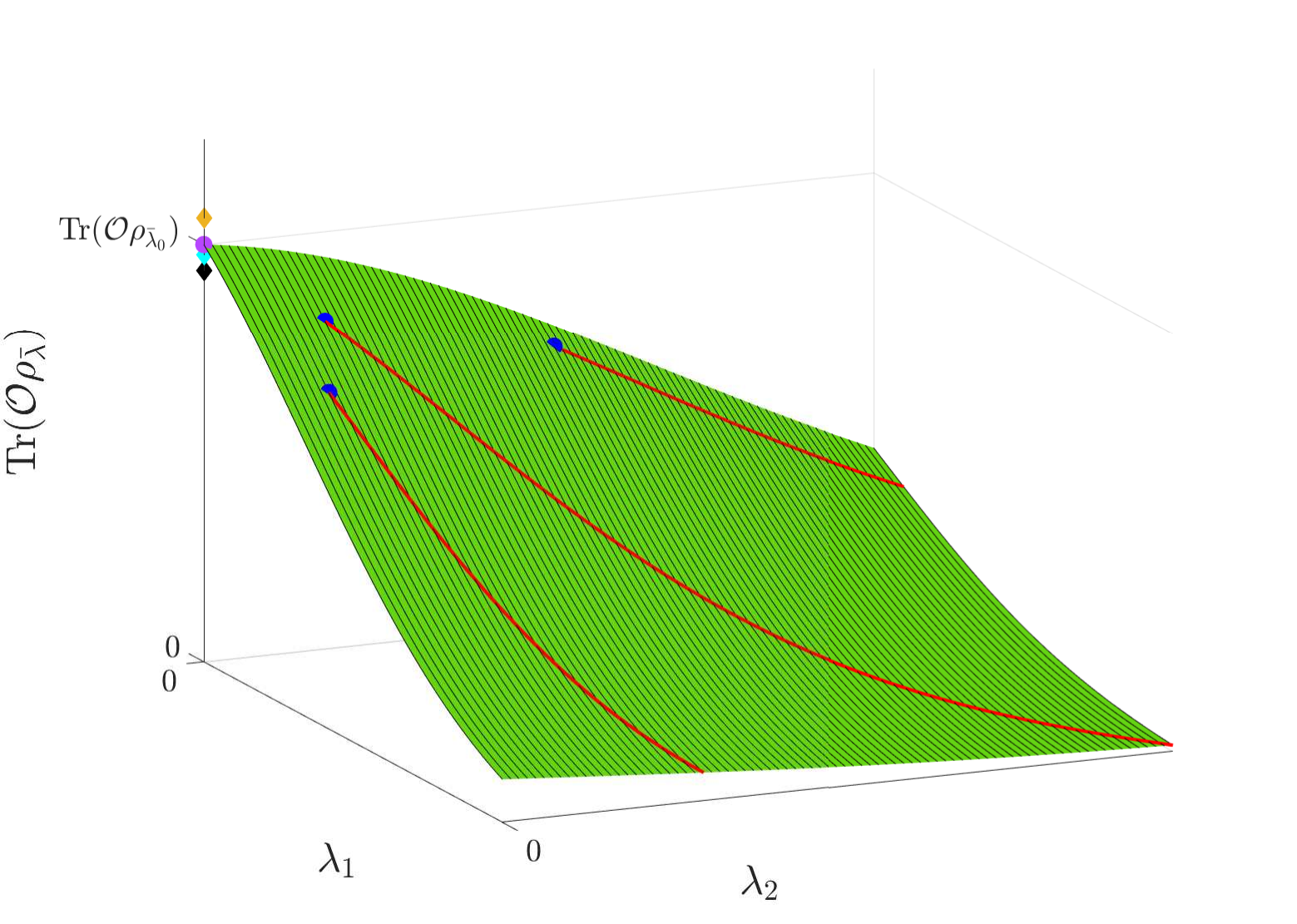}
    \caption{Example of how amplifying all noise sources with a constant factor $G$ reduces the surface related to the multinomial to one variable trajectories, polynomials. The green surface represents the noisy observables as a multivariate function (in the infinite sampling regime). The red lines represent the one dimensional reduction from amplifying all sources by $G$. The blue dots represent different base noise sets $\bar{\lambda}_1$, and the establish the possible polynomials. The purple dot represents the ideal expectation value. The black, light blue and yellow diamonds represent a depiction of the recovered values from each of the trajectories.}
    \label{fig:amps}
\end{figure}

The idea of ZNE was introduced in \cite{TemmeZNE} and experimentally implemented in \cite{ZNEexp1}, where the authors discussed that the evolution of an open system with initial state, $\rho_0$, is given by
\begin{align}\label{eq:LindbladTeme}
    \frac{\partial}{\partial t} \rho(t) = -\frac{i}{\hbar}[H(t),\rho(t)] + \lambda \mathcal{L}(\rho(t)),
\end{align}
where the noise parameter is required to be $\lambda\ll1$ indicating a weak action of the noise. Note that the noise here represents dominant incoherent errors as a result of coupling to the environment \footnote{Coherent errors can also be mitigated using this technique and actually can be transformed into Pauli noise by means of twirling \cite{QEMreview}. Thus, we stick to the noise defined that way for simplicity.}. In a practical setting, there would also be measurement errors that are mitigated by other means \cite{QEMreview}. In ZNE, the aim is to recover the ideal expectation value of a target observable, $\Tr(\Obs\rho_{\lambda_0})$, by means of a collection of noisy expectation values $\{\Tr(\Obs\rho_{\lambda_1}),\cdots,\Tr(\Obs\rho_{\lambda_{n+1}})\}$. The authors discuss how to get the amplified noise rates by stretching the pulses used to implement the desired Hamiltonian evolution. The noisy observable is then approximated as a polynomial of order $n$ over $\lambda$ so that the ideal value can be approximated as 
\begin{equation}\label{eq:estimate}
    \Tr(\Obs\rho_{\lambda_0}) \approx \sum_{j=1}^{n+1 }\gamma_j\Tr(\Obs\rho_{\lambda_j}),
\end{equation}
with the coefficients obtained by the Lagrange interpolators, i.e. $\gamma_j = \prod_{m\neq j}\frac{\lambda_m}{\lambda_j-\lambda_m}$. Importantly, if the set of amplified noise rates is written as $\lambda_j = G_j\lambda_1$, where the $G_j$ are seen as the levels of amplification, we can rewrite the coefficients as $\gamma_j = \prod_{m\neq j}\frac{G_m\lambda_1}{G_j\lambda_1-G_m\lambda_1}=\prod_{m\neq j}\frac{G_m}{G_j-G_m}$. This is referred as Richardson extrapolation \cite{TemmeZNE,ZNEexp1,ZNEcheby,ZNEtiron,QEMreview,QEMunify,Czarnik2021errormitigation}. The reason to choose the Lagrange interpolation coefficients comes from the fact that one can do the following polynomial expansion of the expectation value of the noisy observable \cite{TemmeZNE}
\begin{align}\label{eq:polytemme}
    \Tr(\Obs\rho_{\lambda_j}) = \Tr(\Obs\rho_{\lambda_0}) + \sum_{k=1}^{n}a_k G_j^k\lambda_1^k + R(G_j\lambda_1,\mathcal{L},T),
\end{align}
where $R(G_j\lambda_1,\mathcal{L},T)$ is the residual, which is a function of the noise rate $G_j\lambda_1$, the dissipators $\mathcal{L}$ and the time for which the system is let evolve $T$. In a practical scenario, this expectation value is sampled and, thus, a sampling error $\delta$ is introduced \cite{TemmeZNE}. We neglect this term for simplicity. The coefficients $a_k$ depend on the dissipators, the Hamiltonian and the initial density matrix. If the previously explained Richardson extrapolation is done, the result is obtained by plugging \eqref{eq:polytemme} in \eqref{eq:estimate} \cite{TemmeZNE}
\begin{align*}
    \Tr(\Obs\rho_{\lambda_0}) &\approx \Tr(\Obs\rho_{\lambda_0})\left(\sum_{j=1}^{n+1} \gamma_j\right) \\&+ \sum_{k=1}^{n}a_k \lambda_1^k \left(\sum_{j=1}^{n+1}\gamma_jG_j^k\right)+\left(\sum_{j=1}^{n+1} \gamma_j R(G_j\lambda,\mathcal{L},T)\right)\\& = \Tr(\Obs\rho_{\lambda_0}) + \left(\sum_{j=1}^{n+1} \gamma_j R(G_j\lambda,\mathcal{L},T)\right),
\end{align*}
and since the coefficients fulfill $\sum_{j=1}^{n+1}\gamma_j = 1$ and $\sum_{j=1}^{n+1}\gamma_jG_j^k = 0$, terms up to order $n$ in the polynomial expansion in equation \eqref{eq:polytemme} are cancelled. It is important to see that by doing this the function that is being fitted is a function of the amplifications $G$ and not of the noise rate. The base noise rate, $\lambda_1$, will determine the bias that the estimate will have, $\mathcal{O}(\lambda_1)$ \cite{TemmeZNE}. Following this logic, a similar approach can be taken for the setting in equation \eqref{eq:inidLindblad}. The noisy expectation value can be expressed as a multinomial expansion
\begin{align}\nonumber\label{eq:multinomialexpansion}
     \Tr(\Obs\rho_{\bar{\lambda}}) &= \Tr(\Obs\rho_{\bar{\lambda}_0}) + \sum_{k}a_k \lambda_k +\\& +\sum_k\sum_l b_{kl}\lambda_k\lambda_l + \sum_k\sum_l\sum_m c_{klm}\lambda_k\lambda_l\lambda_m + \cdots,
\end{align}
where we define $\bar{\lambda} = (\lambda_1,\cdots,\lambda_N)$, with $N$ the total amount of possible dissipators in equation \eqref{eq:inidLindblad} and parameters $a_k,b_{kl},c_{klm},\cdots$ are functions of the dissipators, the Hamiltonian and the initial density matrix; similar to the coefficients in \eqref{eq:polytemme}. We have omitted the residual terms and sampling errors for simplicity. Note that this is equivalent to the multinomial expression obtained in \cite{otten}. As noted by Otten and Gray, for a given truncation order of the multinomial, $n$, and $N$ noise terms, the number of coefficients equals to $\sum_{i=0}^n \binom{i+N-1}{N-1}$, which leads to an enormous number of coefficients and leads to the huge measurement overhead required for the ZNE method in \cite{otten}. However, note that if we amplify the noise rates of the system in a uniform manner, i.e. $\bar{\lambda}_j = G_j\bar{\lambda}_1$ with $\bar{\lambda}_1 = (\lambda_1^1,\cdots,\lambda_N^1)$, which means that every rate of each of the noise source is incremented by such factor, then the multinomial expression in \eqref{eq:multinomialexpansion} can be written as
\begin{align*}\label{eq:multinomialexpansion2}
     \Tr(\Obs\rho_{\bar{\lambda}_j}) &= \Tr(\Obs\rho_{\bar{\lambda}_0}) + G_j\left(\sum_{k}a_k \lambda_k^1\right) +\\& +G_j^2\left(\sum_k\sum_l b_{kl}\lambda_k^1\lambda_l^1\right) \\&+ G_j^3\left(\sum_k\sum_l\sum_m c_{klm}\lambda_k^1\lambda_l^1\lambda_m^1\right) + \cdots,
\end{align*}
and, therefore, it still remains as a single variable polynomial over the free variable $G$. In this way, one can select a set of amplifications $\{G_j\}$ for then aiming to fit the polynomial. Hence, the previously explained Richardson extrapolation can be done over this polynomial, with just the same amount of measurement overhead as for the single noise rate case. Therefore, the term ``tunable global noise source'' does not refer to the fact that there is a single noise rate that defines every interaction in the system, but it means that every rate of each noise source can be increased by the same proportion.
\begin{definition}[Tunable global noise source]
A tunable global noise source refers to the fact that the rates of each of the noise sources affecting the qubits of the processor can be tuned by the same factor $G$.
\end{definition}
In this sense, we can rewrite \eqref{eq:inidLindblad} compactly as
\begin{align}\nonumber
       \frac{\partial}{\partial t} \rho(t) &= -\frac{i}{\hbar}[H(t),\rho(t)] + G\left(\sum_{i} \lambda_i \mathcal{L}_i(\rho(t))\right)\\ & = -\frac{i}{\hbar}[H(t),\rho(t)] + G\mathcal{L}(\rho(t)),
\end{align}
where this resembles equation \eqref{eq:LindbladTeme}. The main difference with the picture provided in the ZNE literature \cite{TemmeZNE,ZNEexp1,ZNEcheby,ZNEtiron,QEMreview,QEMunify,Czarnik2021errormitigation} is that the individual terms that make the full dissipator $\mathcal{L}$ now include their own rates. The vector of rates $\bar{\lambda}$ should fulfill the fact that the rates are much smaller than 1 indicating a weak action of the noise and its base level will relate to the actual bias of the estimator. We think that this redefinition makes clearer what the requirement of ``tunable global noise source'' refers to in this context. Figure \ref{fig:amps} shows a graphical example of the reduction of the multivariate function to a polynomial over $G$.

A possible question could be how such global amplification for all the noise sources can be performed. Note that the pulse stretching amplification protocol in \cite{TemmeZNE} and experimentally demonstrated in \cite{ZNEexp1} does indeed obtain this feature \footnote{The device in \cite{ZNEexp1} had a non-uniform error rate distribution. As shown here, the used pulse stretching protocol does obtain the required uniform scaling for such setting.}. The protocol stretches the Hamiltonian time as $T\rightarrow T'=GT$ and rescales the Hamiltonian drive as $H(t)\rightarrow H'(t)=G^{-1}H(G^{-1}t)$, which also implies $\rho(t)\rightarrow\rho'(t)=\rho(G^{-1}t)$. Then solving \eqref{eq:inidLindblad}

\begin{widetext}
\begin{align*}
    \rho'_{\bar{\lambda}}(GT) &= \rho(t=0)-\frac{i}{\hbar}\int_0^{GT}[H'(t),\rho'(t)] dt + \sum_i \lambda_i \int_0^{GT}\mathcal{L}_i(\rho'(t))dt \\& = \rho(t'=0) - \frac{i}{\hbar}\int_0^{T} [G^{-1}H(t'),\rho(t')]G dt' + \sum_i \lambda_i\int_0^{T} \mathcal{L}_i(\rho(t')) G dt' \\& = \rho(t=0) - \frac{i}{\hbar}\int_0^{T} [H(t'),\rho(t')]dt' + \sum_i G \lambda_i\int_0^{T} \mathcal{L}_i(\rho(t')) dt' \\& = \rho_{G\bar{\lambda}}(T),
\end{align*}
\end{widetext}
where in the second equality the change of variable $t=Gt'\rightarrow dt=Gdt'$ has been applied. Therefore, the pulse stretching method achieves what we require. It is difficult to say if the authors for the original proposal for ZNE in \cite{TemmeZNE} had this definition of tunable global noise source in mind, specially since the dissipator in their master equation may include a weighted sum of operators. However, their definition of the $\lambda$ as the noise rate makes the definition somehow confusing and we consider important to clarify the meaning of a tunable global noise source as done here. In fact, the definition given in \cite{TemmeZNE} could be problematic if the generator is defined with the non-uniform couplings embedded, i.e. $\mathcal{L}=\sum_i\lambda'_i\mathcal{L}_i$, then the constraint of weak action of the noise can be violated even if the condition on the global $\lambda<<1$ is set. Our definition imposes the weak action condition on each of the noise sources in the generator avoiding this problematic because the amplification parameter is actually ``artificial'' and not a coupling rate. Other way of doing this proportional increase of all the noise rates of the quantum processor in the system is by means of the so-called Probabilistic Error Amplification (PEA) \cite{ZNEevidence}, which is more convenient considering that pulse level control is not always accessible. This method includes the amplification factor for each of the noise source by random insertion of gates in the circuit. This comes from the fact that circuit must be Pauli twirled \cite{ZNEevidence,tvqc} so that the noise is described by stochastic Pauli errors. Note that coherent errors also transform into stochastic noise due to the twirling process \cite{QEMreview}. The noise rates must be efficiently learned so that each of the noise sources can be amplified the considered factors $G_j$ by appropriate tunings of the probabilities in which those extra Pauli gates must be inserted in each location of the processor. The probabilistic nature of such amplification protocol requires to be careful in the number of circuit runs to be executed so that the obtained events represent the new probabilities of those events in the finite sampling regime. The authors of \cite{ZNEevidence} approximate the noise to an sparse Lindblad-Pauli model so that the learning can be done in an efficient manner. The fact that the multidimensional nature of the such noise model is reduced to a single parameter $G$ can be seen in Figure 2b of \cite{ZNEevidence}. Note that these amplification methods can be applied in any quantum processor. This comes from the fact that the method is based on twirling the noise channels (involves Pauli gates), learning the noise rates and amplifying the noise by probabilistic Pauli gate insertion. It is contradictory that in a gate based quantum computer the insertion of single qubit gates is forbidden. The noise can also be effectively learnt as shown in \cite{ZNEevidence} and note that Otten and Gray do require to learn the coupling rates for their hypersurface method \cite{otten}. Thus, PEA can always be applied in the context of quantum computing and, more specifically, in the scenario discussed in \cite{otten}. Also, other extrapolation methods such as exponential extrapolation can be also applied over the amplification parameter $G$ \cite{ZNEexp1}.

\section{Overhead reduction of extrapolation}
As explained in the previous section, the multinomial expansion presented in \eqref{eq:multinomialexpansion} consists of $\sum_{i=0}^n \binom{i+N-1}{N-1}$ coefficients for a truncation order $n$ and $N$ noise rates. For the method of \cite{otten}, this implies the same number of unknowns required to solve the least square problem they pose to approximate the multinomial whose intercept would be the zero noise estimate of the target observable. Thus, the method requires a measurement overhead equal to such unknown number where each of the measurement is done for a different set of noise rates. To put up some number to the humongous overhead required, let us consider a $100$-qubit processor, where each of the qubits suffer amplitude and dephasing independently (we discard correlated noise terms for simplicity). 
The size of this processor is state-of-the-art \cite{ZNEevidence}. Therefore, in such T1/T2 model, there would be $N=200$ noise rates. We also set the truncation order in $n=3$, which is a fair enough approximation order. The method in \cite{otten} would require $1353400$ measurements of noisy expectation values to work, where each of those are obtained by sampling the machine repeatedly. In \cite{otten}, the authors claim that, for example, they could use the fluctuations of the noise parameters \cite{tvqc,ftvqc} to get those samples at different noise rate sets, at least in superconducting qubit platforms. The parameters of a superconducting qubit are stable for an order of minutes, implying that by this means one would need 2.5 years approximately to get all the required samples. In \cite{otten}, the authors also say that one can use measurements of different processors to get those samples, but for this example more than $10^6$ processors would be required. Both those scenarios are impractical for quantum computing purposes. Note also that in this overhead we have neglected the overhead required to estimate the actual noise rates for the system, needed to execute the protocol proposed in \cite{otten}.

Nevertheless, a correct definition of what a tunable global noise source means, the estimation of the expectation value of the target observable can be obtained by means of $4$ noisy expectation values for the same truncation order. This can be done in a single stability period of a single quantum processor as experimentally shown in \cite{ZNEevidence}, implying that ZNE can be done in a practical manner even in the setting discussed by Otten and Gray, at least in the context of computing.

\section{Conclusion}
In this comment, we have shown that standard ZNE techniques can be applied for the scenario discussed by Otten and Gray in \cite{otten} considering multiqubit noise with arbitrary noise sources and rates so that estimates of the ideal expectation values of a target observable are recovered in the context of computing. As a result, we show that, for such systems, the measurement overhead is reduced to practical values when compared to their proposed method. For doing so, we have thoroughly defined what is meant by ``tunable global noise source'' in the context of ZNE. While many people in the community might be aware of what is meant by such requirement, we consider that it is important to explicitly state it as done here so that everyone in the community correctly understands it as the usual literature might lead to confusion \cite{TemmeZNE,ZNEexp1,ZNEcheby,ZNEtiron,QEMreview,QEMunify,Czarnik2021errormitigation}. Note, for example, that in \cite{ZNEevidence}, the authors are implicitly using this fact for their ZNE experiments, but they rely on the typical definitions of ZNE. As demonstrated in this comment, misunderstanding it might lead to confussion regarding which scenarios standard ZNE can be successful to estimate noise-free observables.

Note, however, that the method in \cite{otten} is still valid for error mitigation and it may find applications in quantum sensing as discussed by the authors. In such setting, it is unclear how the global amplification can be done and, therefore, one might rely on clusters of quantum sensors for applying Otten and Gray's mitigation method.

Finally, one can think when the method proposed by Otten and Gray may lead to more accurate estimates of the target observable. Note that by using tunable global noise source, the best achievable bias will be related to the initial set of noise rates, $\bar{\lambda}_1$, depicted by the blue dots in figure \ref{fig:amps}. The method in \cite{otten} considers more values in the possible parameter space and, therefore, might be more accurate. This can be seen as future work. However, those future studies on the topic should consider the impractical measurement overhead of the method, i.e. such should be reduced by, for example, sparsifying the feature matrix by somehow discriminating which of the noise rates/sources are the most important. This is important since their standard method seems impractical for quantum computing applications in its current form.

\section*{Acknowledgements}
This work was
supported by the Spanish Ministry of Science and Innovation through the project ``Few-qubit quantum hardware, algorithms and codes, on photonic and solid-state systems'' (PLEC2021-008251), by the Diputaci\'on Foral de Gipuzkoa through the ``Quantum error mitigation for near-term quantum computers'' project (IS172551023) and by the Spanish Ministry of Economy and Competitiveness through the MADDIE project (PID2022-137099NB-C44).

\bibliography{bibliography}

\begin{thebibliography}{16}%
\makeatletter
\providecommand \@ifxundefined [1]{%
 \@ifx{#1\undefined}
}%
\providecommand \@ifnum [1]{%
 \ifnum #1\expandafter \@firstoftwo
 \else \expandafter \@secondoftwo
 \fi
}%
\providecommand \@ifx [1]{%
 \ifx #1\expandafter \@firstoftwo
 \else \expandafter \@secondoftwo
 \fi
}%
\providecommand \natexlab [1]{#1}%
\providecommand \enquote  [1]{``#1''}%
\providecommand \bibnamefont  [1]{#1}%
\providecommand \bibfnamefont [1]{#1}%
\providecommand \citenamefont [1]{#1}%
\providecommand \href@noop [0]{\@secondoftwo}%
\providecommand \href [0]{\begingroup \@sanitize@url \@href}%
\providecommand \@href[1]{\@@startlink{#1}\@@href}%
\providecommand \@@href[1]{\endgroup#1\@@endlink}%
\providecommand \@sanitize@url [0]{\catcode `\\12\catcode `\$12\catcode `\&12\catcode `\#12\catcode `\^12\catcode `\_12\catcode `\%12\relax}%
\providecommand \@@startlink[1]{}%
\providecommand \@@endlink[0]{}%
\providecommand \url  [0]{\begingroup\@sanitize@url \@url }%
\providecommand \@url [1]{\endgroup\@href {#1}{\urlprefix }}%
\providecommand \urlprefix  [0]{URL }%
\providecommand \Eprint [0]{\href }%
\providecommand \doibase [0]{http://dx.doi.org/}%
\providecommand \selectlanguage [0]{\@gobble}%
\providecommand \bibinfo  [0]{\@secondoftwo}%
\providecommand \bibfield  [0]{\@secondoftwo}%
\providecommand \translation [1]{[#1]}%
\providecommand \BibitemOpen [0]{}%
\providecommand \bibitemStop [0]{}%
\providecommand \bibitemNoStop [0]{.\EOS\space}%
\providecommand \EOS [0]{\spacefactor3000\relax}%
\providecommand \BibitemShut  [1]{\csname bibitem#1\endcsname}%
\let\auto@bib@innerbib\@empty
\bibitem [{\citenamefont {Otten}\ and\ \citenamefont {Gray}(2019)}]{otten}%
  \BibitemOpen
  \bibfield  {author} {\bibinfo {author} {\bibfnamefont {M.}~\bibnamefont {Otten}}\ and\ \bibinfo {author} {\bibfnamefont {S.~K.}\ \bibnamefont {Gray}},\ }\href {\doibase 10.1103/PhysRevA.99.012338} {\bibfield  {journal} {\bibinfo  {journal} {Phys. Rev. A}\ }\textbf {\bibinfo {volume} {99}},\ \bibinfo {pages} {012338} (\bibinfo {year} {2019})}\BibitemShut {NoStop}%
\bibitem [{\citenamefont {Temme}\ \emph {et~al.}(2017)\citenamefont {Temme}, \citenamefont {Bravyi},\ and\ \citenamefont {Gambetta}}]{TemmeZNE}%
  \BibitemOpen
  \bibfield  {author} {\bibinfo {author} {\bibfnamefont {K.}~\bibnamefont {Temme}}, \bibinfo {author} {\bibfnamefont {S.}~\bibnamefont {Bravyi}}, \ and\ \bibinfo {author} {\bibfnamefont {J.~M.}\ \bibnamefont {Gambetta}},\ }\href {\doibase 10.1103/PhysRevLett.119.180509} {\bibfield  {journal} {\bibinfo  {journal} {Phys. Rev. Lett.}\ }\textbf {\bibinfo {volume} {119}},\ \bibinfo {pages} {180509} (\bibinfo {year} {2017})}\BibitemShut {NoStop}%
\bibitem [{Note1()}]{Note1}%
  \BibitemOpen
  \bibinfo {note} {$Z$ refers to the Pauli Z matrix.}\BibitemShut {Stop}%
\bibitem [{\citenamefont {{deMarti iOlius}}\ \emph {et~al.}(2022)\citenamefont {{deMarti iOlius}}, \citenamefont {{Etxezarreta Martinez}}, \citenamefont {Fuentes}, \citenamefont {Crespo},\ and\ \citenamefont {Garcia-Frias}}]{ton1}%
  \BibitemOpen
  \bibfield  {author} {\bibinfo {author} {\bibfnamefont {A.}~\bibnamefont {{deMarti iOlius}}}, \bibinfo {author} {\bibfnamefont {J.}~\bibnamefont {{Etxezarreta Martinez}}}, \bibinfo {author} {\bibfnamefont {P.}~\bibnamefont {Fuentes}}, \bibinfo {author} {\bibfnamefont {P.~M.}\ \bibnamefont {Crespo}}, \ and\ \bibinfo {author} {\bibfnamefont {J.}~\bibnamefont {Garcia-Frias}},\ }\href {\doibase 10.1103/PhysRevA.106.062428} {\bibfield  {journal} {\bibinfo  {journal} {Phys. Rev. A}\ }\textbf {\bibinfo {volume} {106}},\ \bibinfo {pages} {062428} (\bibinfo {year} {2022})}\BibitemShut {NoStop}%
\bibitem [{\citenamefont {{deMarti iOlius}}\ \emph {et~al.}(2023)\citenamefont {{deMarti iOlius}}, \citenamefont {{Etxezarreta Martinez}}, \citenamefont {Fuentes},\ and\ \citenamefont {Crespo}}]{ton2}%
  \BibitemOpen
  \bibfield  {author} {\bibinfo {author} {\bibfnamefont {A.}~\bibnamefont {{deMarti iOlius}}}, \bibinfo {author} {\bibfnamefont {J.}~\bibnamefont {{Etxezarreta Martinez}}}, \bibinfo {author} {\bibfnamefont {P.}~\bibnamefont {Fuentes}}, \ and\ \bibinfo {author} {\bibfnamefont {P.~M.}\ \bibnamefont {Crespo}},\ }\href {\doibase 10.1103/PhysRevA.108.022401} {\bibfield  {journal} {\bibinfo  {journal} {Phys. Rev. A}\ }\textbf {\bibinfo {volume} {108}},\ \bibinfo {pages} {022401} (\bibinfo {year} {2023})}\BibitemShut {NoStop}%
\bibitem [{\citenamefont {Kandala}\ \emph {et~al.}(2019)\citenamefont {Kandala}, \citenamefont {Temme}, \citenamefont {C{\'o}rcoles}, \citenamefont {Mezzacapo}, \citenamefont {Chow},\ and\ \citenamefont {Gambetta}}]{ZNEexp1}%
  \BibitemOpen
  \bibfield  {author} {\bibinfo {author} {\bibfnamefont {A.}~\bibnamefont {Kandala}}, \bibinfo {author} {\bibfnamefont {K.}~\bibnamefont {Temme}}, \bibinfo {author} {\bibfnamefont {A.~D.}\ \bibnamefont {C{\'o}rcoles}}, \bibinfo {author} {\bibfnamefont {A.}~\bibnamefont {Mezzacapo}}, \bibinfo {author} {\bibfnamefont {J.~M.}\ \bibnamefont {Chow}}, \ and\ \bibinfo {author} {\bibfnamefont {J.~M.}\ \bibnamefont {Gambetta}},\ }\href {\doibase 10.1038/s41586-019-1040-7} {\bibfield  {journal} {\bibinfo  {journal} {Nature}\ }\textbf {\bibinfo {volume} {567}},\ \bibinfo {pages} {491} (\bibinfo {year} {2019})}\BibitemShut {NoStop}%
\bibitem [{\citenamefont {Krebsbach}\ \emph {et~al.}(2022)\citenamefont {Krebsbach}, \citenamefont {Trauzettel},\ and\ \citenamefont {Calzona}}]{ZNEcheby}%
  \BibitemOpen
  \bibfield  {author} {\bibinfo {author} {\bibfnamefont {M.}~\bibnamefont {Krebsbach}}, \bibinfo {author} {\bibfnamefont {B.}~\bibnamefont {Trauzettel}}, \ and\ \bibinfo {author} {\bibfnamefont {A.}~\bibnamefont {Calzona}},\ }\href {\doibase 10.1103/PhysRevA.106.062436} {\bibfield  {journal} {\bibinfo  {journal} {Phys. Rev. A}\ }\textbf {\bibinfo {volume} {106}},\ \bibinfo {pages} {062436} (\bibinfo {year} {2022})}\BibitemShut {NoStop}%
\bibitem [{\citenamefont {Giurgica-Tiron}\ \emph {et~al.}(2020)\citenamefont {Giurgica-Tiron}, \citenamefont {Hindy}, \citenamefont {LaRose}, \citenamefont {Mari},\ and\ \citenamefont {Zeng}}]{ZNEtiron}%
  \BibitemOpen
  \bibfield  {author} {\bibinfo {author} {\bibfnamefont {T.}~\bibnamefont {Giurgica-Tiron}}, \bibinfo {author} {\bibfnamefont {Y.}~\bibnamefont {Hindy}}, \bibinfo {author} {\bibfnamefont {R.}~\bibnamefont {LaRose}}, \bibinfo {author} {\bibfnamefont {A.}~\bibnamefont {Mari}}, \ and\ \bibinfo {author} {\bibfnamefont {W.~J.}\ \bibnamefont {Zeng}},\ }in\ \href {\doibase 10.1109/QCE49297.2020.00045} {\emph {\bibinfo {booktitle} {2020 IEEE International Conference on Quantum Computing and Engineering (QCE)}}}\ (\bibinfo {year} {2020})\ pp.\ \bibinfo {pages} {306--316}\BibitemShut {NoStop}%
\bibitem [{\citenamefont {Cai}\ \emph {et~al.}(2023)\citenamefont {Cai}, \citenamefont {Babbush}, \citenamefont {Benjamin}, \citenamefont {Endo}, \citenamefont {Huggins}, \citenamefont {Li}, \citenamefont {McClean},\ and\ \citenamefont {O'Brien}}]{QEMreview}%
  \BibitemOpen
  \bibfield  {author} {\bibinfo {author} {\bibfnamefont {Z.}~\bibnamefont {Cai}}, \bibinfo {author} {\bibfnamefont {R.}~\bibnamefont {Babbush}}, \bibinfo {author} {\bibfnamefont {S.~C.}\ \bibnamefont {Benjamin}}, \bibinfo {author} {\bibfnamefont {S.}~\bibnamefont {Endo}}, \bibinfo {author} {\bibfnamefont {W.~J.}\ \bibnamefont {Huggins}}, \bibinfo {author} {\bibfnamefont {Y.}~\bibnamefont {Li}}, \bibinfo {author} {\bibfnamefont {J.~R.}\ \bibnamefont {McClean}}, \ and\ \bibinfo {author} {\bibfnamefont {T.~E.}\ \bibnamefont {O'Brien}},\ }\href {\doibase 10.1103/RevModPhys.95.045005} {\bibfield  {journal} {\bibinfo  {journal} {Rev. Mod. Phys.}\ }\textbf {\bibinfo {volume} {95}},\ \bibinfo {pages} {045005} (\bibinfo {year} {2023})}\BibitemShut {NoStop}%
\bibitem [{\citenamefont {Bultrini}\ \emph {et~al.}(2023)\citenamefont {Bultrini}, \citenamefont {Gordon}, \citenamefont {Czarnik}, \citenamefont {Arrasmith}, \citenamefont {Cerezo}, \citenamefont {Coles},\ and\ \citenamefont {Cincio}}]{QEMunify}%
  \BibitemOpen
  \bibfield  {author} {\bibinfo {author} {\bibfnamefont {D.}~\bibnamefont {Bultrini}}, \bibinfo {author} {\bibfnamefont {M.~H.}\ \bibnamefont {Gordon}}, \bibinfo {author} {\bibfnamefont {P.}~\bibnamefont {Czarnik}}, \bibinfo {author} {\bibfnamefont {A.}~\bibnamefont {Arrasmith}}, \bibinfo {author} {\bibfnamefont {M.}~\bibnamefont {Cerezo}}, \bibinfo {author} {\bibfnamefont {P.~J.}\ \bibnamefont {Coles}}, \ and\ \bibinfo {author} {\bibfnamefont {L.}~\bibnamefont {Cincio}},\ }\href {\doibase 10.22331/q-2023-06-06-1034} {\bibfield  {journal} {\bibinfo  {journal} {{Quantum}}\ }\textbf {\bibinfo {volume} {7}},\ \bibinfo {pages} {1034} (\bibinfo {year} {2023})}\BibitemShut {NoStop}%
\bibitem [{\citenamefont {Czarnik}\ \emph {et~al.}(2021)\citenamefont {Czarnik}, \citenamefont {Arrasmith}, \citenamefont {Coles},\ and\ \citenamefont {Cincio}}]{Czarnik2021errormitigation}%
  \BibitemOpen
  \bibfield  {author} {\bibinfo {author} {\bibfnamefont {P.}~\bibnamefont {Czarnik}}, \bibinfo {author} {\bibfnamefont {A.}~\bibnamefont {Arrasmith}}, \bibinfo {author} {\bibfnamefont {P.~J.}\ \bibnamefont {Coles}}, \ and\ \bibinfo {author} {\bibfnamefont {L.}~\bibnamefont {Cincio}},\ }\href {\doibase 10.22331/q-2021-11-26-592} {\bibfield  {journal} {\bibinfo  {journal} {{Quantum}}\ }\textbf {\bibinfo {volume} {5}},\ \bibinfo {pages} {592} (\bibinfo {year} {2021})}\BibitemShut {NoStop}%
\bibitem [{Note2()}]{Note2}%
  \BibitemOpen
  \bibinfo {note} {Coherent errors can also be mitigated using this technique and actually can be transformed into Pauli noise by means of twirling \cite {QEMreview}. Thus, we stick to the noise defined that way for simplicity.}\BibitemShut {Stop}%
\bibitem [{Note3()}]{Note3}%
  \BibitemOpen
  \bibinfo {note} {The device in \cite {ZNEexp1} had a non-uniform error rate distribution. As shown here, the used pulse stretching protocol does obtain the required uniform scaling for such setting.}\BibitemShut {Stop}%
\bibitem [{\citenamefont {Kim}\ \emph {et~al.}(2023)\citenamefont {Kim}, \citenamefont {Eddins}, \citenamefont {Anand}, \citenamefont {Wei}, \citenamefont {van~den Berg}, \citenamefont {Rosenblatt}, \citenamefont {Nayfeh}, \citenamefont {Wu}, \citenamefont {Zaletel}, \citenamefont {Temme},\ and\ \citenamefont {Kandala}}]{ZNEevidence}%
  \BibitemOpen
  \bibfield  {author} {\bibinfo {author} {\bibfnamefont {Y.}~\bibnamefont {Kim}}, \bibinfo {author} {\bibfnamefont {A.}~\bibnamefont {Eddins}}, \bibinfo {author} {\bibfnamefont {S.}~\bibnamefont {Anand}}, \bibinfo {author} {\bibfnamefont {K.~X.}\ \bibnamefont {Wei}}, \bibinfo {author} {\bibfnamefont {E.}~\bibnamefont {van~den Berg}}, \bibinfo {author} {\bibfnamefont {S.}~\bibnamefont {Rosenblatt}}, \bibinfo {author} {\bibfnamefont {H.}~\bibnamefont {Nayfeh}}, \bibinfo {author} {\bibfnamefont {Y.}~\bibnamefont {Wu}}, \bibinfo {author} {\bibfnamefont {M.}~\bibnamefont {Zaletel}}, \bibinfo {author} {\bibfnamefont {K.}~\bibnamefont {Temme}}, \ and\ \bibinfo {author} {\bibfnamefont {A.}~\bibnamefont {Kandala}},\ }\href {\doibase 10.1038/s41586-023-06096-3} {\bibfield  {journal} {\bibinfo  {journal} {Nature}\ }\textbf {\bibinfo {volume} {618}},\ \bibinfo {pages} {500} (\bibinfo {year} {2023})}\BibitemShut {NoStop}%
\bibitem [{\citenamefont {{Etxezarreta Martinez}}\ \emph {et~al.}(2021)\citenamefont {{Etxezarreta Martinez}}, \citenamefont {Fuentes}, \citenamefont {Crespo},\ and\ \citenamefont {Garcia-Frias}}]{tvqc}%
  \BibitemOpen
  \bibfield  {author} {\bibinfo {author} {\bibfnamefont {J.}~\bibnamefont {{Etxezarreta Martinez}}}, \bibinfo {author} {\bibfnamefont {P.}~\bibnamefont {Fuentes}}, \bibinfo {author} {\bibfnamefont {P.}~\bibnamefont {Crespo}}, \ and\ \bibinfo {author} {\bibfnamefont {J.}~\bibnamefont {Garcia-Frias}},\ }\href {\doibase 10.1038/s41534-021-00448-5} {\bibfield  {journal} {\bibinfo  {journal} {npj Quantum Information}\ }\textbf {\bibinfo {volume} {7}},\ \bibinfo {pages} {115} (\bibinfo {year} {2021})}\BibitemShut {NoStop}%
\bibitem [{\citenamefont {Etxezarreta~Martinez}\ \emph {et~al.}(2023)\citenamefont {Etxezarreta~Martinez}, \citenamefont {Fuentes}, \citenamefont {deMarti iOlius}, \citenamefont {Garcia-Frias}, \citenamefont {Fonollosa},\ and\ \citenamefont {Crespo}}]{ftvqc}%
  \BibitemOpen
  \bibfield  {author} {\bibinfo {author} {\bibfnamefont {J.}~\bibnamefont {Etxezarreta~Martinez}}, \bibinfo {author} {\bibfnamefont {P.}~\bibnamefont {Fuentes}}, \bibinfo {author} {\bibfnamefont {A.}~\bibnamefont {deMarti iOlius}}, \bibinfo {author} {\bibfnamefont {J.}~\bibnamefont {Garcia-Frias}}, \bibinfo {author} {\bibfnamefont {J.~R.}\ \bibnamefont {Fonollosa}}, \ and\ \bibinfo {author} {\bibfnamefont {P.~M.}\ \bibnamefont {Crespo}},\ }\href {\doibase 10.1103/PhysRevResearch.5.033055} {\bibfield  {journal} {\bibinfo  {journal} {Phys. Rev. Res.}\ }\textbf {\bibinfo {volume} {5}},\ \bibinfo {pages} {033055} (\bibinfo {year} {2023})}\BibitemShut {NoStop}%
\end{thebibliography}%

\end{document}